# Non-parametric estimation of Expectation and Variance of event count and of incidence rate in clinical trial - where intensity of event-occurrence changes with the occurrence of each higher order event

## Sudipta Bhattacharya

Abstract: Counting processes where intensity of event occurrence changes with the occurrence of each higher order event are often experienced in clinical trials (e.g. cardio-vascular events) and also in various other scientific fields. There are methods for estimating event rates or mean number of event counts and related parameters. The method proposed in this paper not only generalizes all the existing methods for estimating the mean count of events but also estimates the mean of counts over time and the upper limit of the variance of counts over time of events generated by the above-mentioned event generating process in a completely non-parametric set-up. In addition, the proposed method is applied on simulated data to estimate the mean for that simulated process and those estimates are compared with the Nelson Aalen estimates. Using the estimate for maximum value of variance, asymptotic Normality of subject's event counts and for sampling distribution of mean event counts for a population (i.e., the incidence rate for that population) is established, which can be used for calculating confidence interval of the estimate of mean counts.

1       **Background**

Analysis of recurrent events is an important branch of statistical analysis, especially in the context of clinical trials and epidemiological research. Relapses and recurrences of events like development of new lesion in metastatic cancer, migraine attack, asthma, stoke and cardio-vascular events are common incidents in clinical studies, each of which generates recurrent event data. Example of similar events in social science would be recurrent drug abuse or a specific type of crime recurrently committed by a subject.

In reality, there are situations (e.g., progression of disease through new lesion in cancer, armed robberries) where every time subjects experience a new event, they advance to a new state, where a new intensity function for the successive event occurrence is created; this makes the occurrence of events dependent on the history of the process through preceding events. Progressive diseases like cancer, Parkinson's disease, Cardio-Vascular diseases, etc. or recurrent drug abuse or violent crimes by an offender are examples of such situations.

Mathematically speaking, the definition of intensity function for event occurrences in any point process is as follows:

$$\lambda(t|\mathcal{H}_t) = \lim_{\Delta t \to 0} \frac{\Pr(\Delta X_t = 1|\mathcal{H}_{0 \le s < t})}{\Delta t} = \lim_{\Delta t \to 0} \frac{\Pr(\Delta X_t = 1|\mathcal{H}_t)}{\Delta t},$$

(1.1)

where $X_t = X(t) = X(0, t)$ denotes the count of event occurrences within time interval of $(0, t]$, and $\mathcal{H}_t = \mathcal{H}_{0 \le s < t}$ is the history of the process generating recurrent events by time $t$ (Cook and Lawless [1], p 28).

For a general point process where occurrence of successive events depends on the history of the process especially through preceding events (that is, when intensity of event occurrence changes with the occurrence of each higher order event), the most generalized expression for mean-function of counts <u>over time</u> $t$ in expectation-form can be made using equation (1.1) in the following way:

$$\mu_t = E^X[X_t] = E^{\mathcal{H}}[E^X\{X_t|\mathcal{H}_t\}] = E^{\mathcal{H}}\left[E^X\left\{\int_0^t (dX_u|\mathcal{H}_u)\right\}\right] = E^{\mathcal{H}}\left[\int_0^t E^X\{dX_u|\mathcal{H}_u\}\right] =$$

$$E^{\mathcal{H}}\left[\int_0^t P(dX_u = 1|\mathcal{H}_u)\right] = \int_0^t E^{\mathcal{H}}[P(dX_u = 1|\mathcal{H}_u)] = \int_0^t E^{\mathcal{H}}[\lambda(t|\mathcal{H}_u)du]$$

(1.2)

[ For the sake of notational simpliciry, throughout this paper, $E^{X_s}[f(s, X_s, \mathcal{H}_s, \theta)]$ will be denoted by $E^X[f(s, X_s, \mathcal{H}_s, \theta)]$ and $E^{\mathcal{H}_s}[f(s, X_s, \mathcal{H}_s, \theta)]$ will be denoted by $E^{\mathcal{H}}[f(s, X_s, \mathcal{H}_s, \theta)]$, where $s$ is the time-point, $X_s$ and $\mathcal{H}_s$ are as defined above and $\theta$ is some unknown parameter.]

It is noteworthy that, unless the occurrence of successive events depends on the history of the process especially through preceding events (i.e., if intensity of event occurrence does not change with the occurrence of each higher order event), the usual way of deriving cumulative mean or rate function <u>over time</u> $t$ in case of a counting or point process is $E^X[X_t] = E^X\left[\int_0^t (dX_u|\mathcal{H}_u)\right] = \int_0^t E^X[dX_u|\mathcal{H}_u] = \int_0^t P[dX_u = 1|\mathcal{H}_u] = \int_0^t \lambda(t|\mathcal{H}_u)du$

(1.3)

The variance ($\sigma_t^2$) can be expressed as $\sigma_t^2 = \text{Var}^X(X_t) = E^X(X_t^2) - \{E^X(X_t)\}^2$.

Where $E^X(X_t^2) = E^{\mathcal{H}}\left[E^X\left\{\int_0^t (dX_u|\mathcal{H}_u)\right\}^2\right] = E^{\mathcal{H}}\left[E^X\{(\int_0^t (dX_u|\mathcal{H}_u)^2) + (\int_0^t \int_0^t (dX_v dX|\mathcal{H}_v, \mathcal{H}_u; u \neq v))\}\right]$ in case of a general counting process, especially where occurrence of successive events depends on history through preceding events.

(1.4)

If $E^{\mathcal{H}}\left[E^X\{\int_0^t \int_0^t (dX_v dX|\mathcal{H}_v, \mathcal{H}_u; u \neq v)\}\right]$ in equation (1.4) is difficult to derive, then the derivation of the variance function may not be possible in case of an event process, where intensity of event occurrence changes with the occurrence of each higher order event.

By definition, the mean number of counts <u>at time</u> t, that is, the cross-sectional mean number of counts, $m_t = \sum_{j=1}^{J} j \times P[X_t = j] = \sum_{j=1}^{J} j \times P_t^{0j}$;

(1.5)

where $J$ corresponds to the maximum number of events occurred to a subject and $P_t^{0j}$ corresponds to the marginal (i.e., not conditional on history of the process) probability of $j^{th}$ event occurrence <u>at time</u> $t$.

In addition, $Var(\widehat{m_t}) = Var\left[\sum_{j=1}^{J} j \times \widehat{P_t^{0j}}\right]$

$$= \sum_{j=1}^{J} j^2 \times Var\left[\widehat{P_t^{0j}}\right] + \sum_{j=1}^{J} \sum_{l \neq j; l=1}^{J} j \times l \times Cov(\widehat{P_t^{0j}}, \widehat{P_t^{0l}})$$

(1.6)

If $Cov(\widehat{P_t^{0j}}, \widehat{P_t^{0l}})$ in equation (1.6) is difficult to derive and estimate, the derivation of variance of the estimator for mean count of events in a process may not be possible.

Nelson-Aalen estimator (Aalen [2]) is a non-parametric estimator of mean function for recurrent processes, which is also an unbiased estimator for mean function when the recurrent events follow Poisson process (Cook and Lawless [1], p 68). The Nelson-Aalen estimate for mean function is also considered to be unbiaseed regardless of the process (Cook and Lawless [1], p 83). Cook, Lawless, Lakhal-Chaieb et. al. [3] also proposed robust estimation methods for mean functions and treatment effects for recurrent event under event dependent censoring and termination. It is noteworthy that the Nelson-Aalen estimating method and also all the methods discussed by Cook, Lawless, Lakhal-Chaieb et. al. [3] for estimating the mean or rate function <u>over time</u> $t$ in case of a counting or point process are implicitly based on assumption described in equation (1.3).

Maller et al. [4] discussed method of estimating cumulative probabilities of higher order event occurrences $F_t^{0j} = P[X_t \geq j]$ using Kaplan-Meier survival probability estimates (Kaplan and Meier [5]); and thereafter used the definition of mean function presented in equation (1.5) for the estimation of mean count of events <u>at time</u> $t$. It is noteworthy that Kaplan-Meier estimates are based on assumption described in equation (1.3).

Dong et. al. [6] also used Kaplan-Meier survival probability estimates of higher order event occurrences for estimating mean cumulative count function. The methodology implicitly adapts the definition of cumulative mean function <u>over time</u> $t$, and is based on assumption described in equation (1.3).

On the same note, all the other different methods introduced by Nelson [7], Doganaksoy and Nelson [8], and Lawless and Nadeau [9] also estimate the mean <u>over time</u> $t$, and are all based on assumption described in equation (1.3).

There are methods available for estimating mean or rate function of recurrent events <u>over time</u>, when intensities of event occurrences in a Poisson process depend on the history of the process through time-varying covariates (but naturally not on preceding events). Andersen-Gill model (Andersen and Gill [10]) and also methods introduced by Lin, Wei and Yang [11], by Lin, Wei and Ying [12], and by Miloslavsky, Keles, Van der Laan et al. [13] (all of which are based on the definition of intensity function defined in Andersen-Gill model) can all be used in this context.

When occurrence of successive events depends on the history of the process especially through preceding events, Aalen-Johansen estimator (Aalen and Johansen [14]) provides non-parametric estimates of the transition probability matrix, which could be utilized for estimating the marginal probabilities ($P_t^{0j}$; $j$ being the index for the higher order events), and the mean count of events at time $t$ can be estimated based on equation (1.5). However, it should be noted that although there are methods available for deriving variance estimates for the Aalen-Johansen estimates of transition probability matrix, since $Cov(\widehat{P_t^{0j}}, \widehat{P_t^{0l}})$ is difficult to estimate when intensity of event occurrence changes with the occurrence of each higher order event, the derivation of variance of the estimator for mean of event counts using Aalen-Johansen transition probability estimates may not be possible, as indicated in the paragraph following equation (1.6).

The discussion provided above indicates that there are several methods for estimating the mean function over time or at any given time-point for conting processes including general counting process, where event occurrences depend on history through time-varying covariates and/or because of drop-outs or terminating events. However, when occurrence of successive events depends on history through preceding events (i.e., when intensity of successive event occurrence changes with the occurrence of preceding events), there exists no straightforward method that can provide an estimator for the marginal (i.e., not conditional on history of the process) moments and therby the mean of event counts (and also the variance of mean estimate). For the same reason, the analysis of incidence rate of event occurrences in a population (where incidence rate is defined as the sum of all event occurrences in the population divided by the total number of subjects at risk in that population) is not possible when intensity of event occurrence changes with the occurrence of each higher order event. Therefore, there exists unmet need for developing methods to estimate the mean, incidence rate, etc. for recurrent events like consecutive cardio-vascular events to a subject or successive events of drug abuse by a subject, etc. That is, a method needs to be developed for estimating cumulative mean or rate function over time for count-data (and variance of that estimate), which can be used in any counting process, including where occurrence of successive events depends on history through preceding events, especially in absence of any covariates.

Following is an overview of how the development of and the discussion about the non-parametric estimator of incidence rate function are laid out in the next few sections for a general counting process where intensity of event occurrence changes with the occurrence of each higher order event. In Section 2, the mathematical development of the proposed estimator is described for generalized cumulative mean function <u>over time</u> in case of a general counting process where occurrence of successive events depends on history through preceding events. In addition, the properties of that estimator are discussed and the maximum value of the variance function for the count of events over time is also derived. Thereafter, the mathematical condition for asymptotic normality of count of events occurred to an individual subject is stated and the sampling distribution of mean number of event counts across subjects (i.e., for the incidence rate function) is discussed. In section 3, simulated data are analyzed to demonstrate the validity and relevance of the proposed estimator of mean over time of the count of events generated by such a general point process, by comparing the results produced by the proposed estimator to the results produced by Nelson-Aalen estimator, which is a non-parametric and unbiased estimator of mean function of recurrent events, under the assumption of Poisson process (and more generally, under criteria set by equation (1.3)). In section 4, the application of the proposed methodology through the use of incidence rate function is discussed, and some further possible extensions of the present work are mentioned.

## 2  Mathematical Development

### 2.1  Mean of event counts

Following equation (1.2), the mean of event counts in a general counting process (especially where intensities of successive event occurances depend on the history of the process through preceding events) can be expressed as:

$$\mu_t = E^X[X_t] = \int_0^t E^{\mathcal{H}}[P(dX_u = 1|\mathcal{H}_u)] = \int_0^t E^{\mathcal{H}}[P(dX_u = 1|\mathcal{H}_{0 \leq s < u})]$$

$$= \sum_{u=0}^{u=t-\Delta t} \{\sum_{\Omega_{\mathcal{H}_{0 \leq s \leq u}}} P(\Delta X_u = 1|\mathcal{H}_{0 \leq s \leq u}) P[\mathcal{H}_{0 \leq s \leq u}]\}$$

Where $\Omega_{\mathcal{H}_{0 \leq s \leq u}} = \Omega_{\mathcal{H}_u}$ is the sigma field generated by $\mathcal{H}_{0 \leq s \leq u} = \mathcal{H}_u$

(2.1.1)

$= \sum_{u=0}^{u=t-\Delta t} \{\sum_{\Omega_{\mathcal{H}_{0 \leq s \leq u}}} (\chi(u|\mathcal{H}_u) \Delta u) P[\mathcal{H}_{0 \leq s \leq u}]\}$, using equation (1.1) and assuming $\lambda_k(u|\mathcal{H}_u^k) = \lambda(u|\mathcal{H}_u)$, where $\lambda_k(u|\mathcal{H}_u^k)$ is the individual intensity function and $\mathcal{H}_u^k$ individual history until time $u$ for subject $k$ and $\lambda(u|\mathcal{H}_u)$ is the population intensity function with population history $\mathcal{H}_u$ until time $u$.

There is no straight-forward way to estimate $P[\mathcal{H}_{0 \leq s < u}]$, especially when occurrence of successive events depends on history through preceding events. A non-parametric approach for estimating the history of the process generating recurrent events is adapted from Menjoge [15], which is described in the following few paragraphs.

For any $u = t_m$, (where $0 < t_1 \leq t_2 \leq \cdots.. \leq t_m \leq \cdots. \leq t_{N_t} \leq t$ are the time-points of event occurrences)

- if $\mathcal{H}_{t_m}$ denotes the history of the occurrence of $m$ events in the entire population (consisting of $n$ subjects) by time $t_m$,
- if $J_{t_m}$ denotes the highest / maximum number of events that occurred to subject(s) in that population by time $t_m$,

- and if $\{H_{t_m}: N = i, i = 1,2,\ldots, J_{t_m}\}$ denote the histories of the events occurred to a monotone sequence of subsets of the population under study over the time period of $0 < u \leq t_m$, then for any $u = t_m$ since the event histories $\{H_{t_m}: N = j\}, j \geq 1$ form a monotone sequence (i.e., since $(H_{t_m}: N = j) = (H_{t_m}: N = j\ only) \subseteq (H_{t_m}: N = \overline{j-1}) = (H_{t_m}: N = j\ only) + (H_{t_m}: N = \overline{j-1}\ only) \subseteq (H_{t_m}: N = \overline{j-2}) = (H_{t_m}: N = j\ only) + (H_{t_m}: N = \overline{j-1}\ only) + (H_{t_m}: N = \overline{j-2}\ only) \subseteq \cdots \subseteq (H_{t_m}: N = 3) \subseteq (H_{t_m}: N = 2) \subseteq (H_{t_m}: N = 1)$, with $(H_{t_m}: N = 0) + (H_{t_m}: N \geq 1) = (H_{t_m}: N = 0) + (H_{t_m}: N = 1) = \mathcal{H}_{t_m}$), hence, $P(\mathcal{H}_{t_m}) = P[H_{t_m}: N = 0\ only] + P[\cup_{j\geq 1}\{H_{t_m}: N = j\ only\}] = P[H_{t_m}: N = 0] + P[H_{t_m}: N \geq 1] = P[H_{t_m}: N = 0] + P[H_{t_m}: N = 1] = 1.$

Using equation (2.1.1) and exploiting the nature of the history of the process generating recurrent events, which is a monotone sequence for higher order events ($j \geq 1$), the following estimator for the mean number of counts can be derived.

$$E^X[X_t] = E^{\mathcal{H}}[E^X\{X_t|\mathcal{H}_t\}] = \sum_{u=0}^{u=t-\Delta t} \{\sum_{\Omega_{\mathcal{H}_{0\leq s\leq u}}} P(\Delta X_u = 1|\mathcal{H}_{0\leq s\leq u})P[\mathcal{H}_{0\leq s\leq u}]\}$$

$$= \sum_{t_i=t_1>0}^{t_i=t_{N_t}\leq t} \sum_{\Omega_{\mathcal{H}_{0\leq s\leq t_i}}} P(\Delta X_{t_i} = 1|\mathcal{H}_{0\leq s\leq t_i})P[\mathcal{H}_{0\leq s\leq t_i}]$$

Therefore,

$$\widehat{\mu_t} = \widehat{E[X_t]} = \sum_{t_i=t_1>0}^{t_i=t_{N_t}\leq t} \{\hat{P}(\Delta X_{t_i} = 1|H_{t_i}: N = 0)\hat{P}[H_{t_i}: N = 0]$$

$$+ \sum_{j=1}^{J_{t_i}} \hat{P}(\Delta X_{t_i} = 1|H_{t_i}: N = j\ only)\hat{P}[H_{t_i}: N = j\ only]\}$$

$$= \sum_{j=0}^{J_{t_{N_t}}} \{\sum_{t_i=t_1>0}^{t_i=t_{N_t}\leq t} \hat{P}(\Delta X_{t_i} = 1 | H_{t_i}: N = j \; only) \hat{P}[H_{t_i}: N = j \; only]\},$$

(2.1.2)

Where $\hat{P}[H_{t_i}: N = 0 \; only] = \hat{P}[H_{t_i}: N = 0]$ and $\hat{P}(\Delta X_{t_i} = 1 | H_{t_i}: N = 0 \; only) = \hat{P}(\Delta X_{t_i} = 1 | H_{t_i}: N = 0)$; and the individual probabilities can be estimated using the Kaplan-Meier survival probability estimates in the following way.

$\hat{P}(\Delta X_{t_i} = 1 | H_{t_i}: N = j \; only), \forall j \geq 0$ is the probability estimate for the $\overline{j+1}^{th}$ event to occur at time instant $t_i + \Delta t$ in a sub-population where <u>only</u> the $j^{th}$ event has already occurred by time $t_i$ (that is, the $\overline{j+1}^{th}$ event has not occurred to this sub-population till time $t_i + \Delta t$);

and $\hat{P}[H_{t_i}: N = j \; only], \forall j \geq 1$ denotes the probability estimate for the occurrence of only $j$ events by time $t_i$, which is to be calculated in the following manner.

$\hat{P}[H_{t_i}: N = j \; only] = \hat{P}[\{(H_{t_i}: N \neq \overline{j+1}) \cap (H_{t_i}: N = j)\} | (H_{t_i}: N = j)] \times \hat{P}[H_{t_i}: N = j | H_{t_i}: N = \overline{j-1}] \times \hat{P}[H_{t_i}: N = \overline{j-1} | H_{t_i}: N = \overline{j-2}] \times ... \times \hat{P}[H_{t_i}: N = 3 | H_{t_i}: N = 2] \times \hat{P}[H_{t_i}: N = 2 | H_{t_i}: N = 1] \times \hat{P}[H_{t_i}: N = 1]$; where:

$\hat{P}[\{(H_{t_i}: N \neq \overline{j+1}) \cap (H_{t_i}: N = j)\} | H_{t_i}: N = j], \forall j \geq 1$ is the (Kaplan-Meier type) estimates for the survival probability of the $\overline{j+1}^{th}$ event occurrence by time $t_i$ in a sub-population where the $j^{th}$ event has already occurred, and

$\hat{P}[H_{t_i}: N = j | H_{t_i}: N = \overline{j-1}], \forall j \geq 2$ are the (Kaplan-Meier type) estimates for the probability of the $j^{th}$ event occurrence by time $t_i$ in a sub-population where the $\overline{j-1}^{th}$ event has already occurred.

$\hat{P}[H_{t_i}: N = 0]$ and $\hat{P}[H_{t_i}: N = 1]$ are the usual Kaplan-Meier estimates for survival and failure probabilities of the first event occurrence by $t_i$ in the entire population.

### 2.1.1 Properties

*Comparabality with Nelson-Aalen Estimator.* Regardless of the process, the numerical equality between the estimate based on the proposed estimator and the Nelson-Aalen estimate in case of no drop-outs is evident from the algebra of the methodology adapted for estimating the proposed estimator of generalized mean function of event recurrences over time.

*Estimator of cumulative mean function over time for Poisson process.* Based on expression presented in equation (2.1.1), it is clear that in case of Poisson process (when $\lambda(u|\mathcal{H}_u) = \rho(u)$), the proposed estimator for the expected number of event occurrences over time boils down to cumulative (over time) intensity function. This also establishes the validity of the proposed estimator as a robust estimator of the cumulative mean function over time, regardless of the process.

## 2.2 Variance of event counts

The variance $(\sigma_t^2)$ can be expressed as $\sigma_t^2 = \text{Var}^X(X_t) = E^X(X_t^2) - \{E^X(X_t)\}^2$.

Now, $E^X(X_t^2) = E^{\mathcal{H}}\left[E^X\left\{\int_0^t (dX_u|\mathcal{H}_u)\right\}^2\right]$

$= E^{\mathcal{H}}\left[E^X\left\{(\int_0^t (dX_u|\mathcal{H}_u)^2) + (\int_0^t \int_0^t (dX_v dX_u|\mathcal{H}_v, \mathcal{H}_u; u \neq v))\right\}\right]$

$\leq E^{\mathcal{H}}\left[E^X\left\{\left(\int_0^t (dX_u|\mathcal{H}_u)^2\right) + ((X_t|\mathcal{H}_t) - 1)(\int_0^t (dX_u|\mathcal{H}_u)^2)\right\}\right]$

$= E^{\mathcal{H}}\left[E^X\left\{(X_t|\mathcal{H}_t)(\int_0^t (dX_u|\mathcal{H}_u)^2)\right\}\right]$

$$= E^{\mathcal{H}}\left[E^X\left\{\int_0^t x_t^{obs}\,((dX_u|\mathcal{H}_u)^2)\right\}\right]$$

$$= x_t^{obs} E^{\mathcal{H}}\left[\int_0^t E^X\{(dX_u|\mathcal{H}_u)^2\}\right]$$

$$= x_t^{obs} E^{\mathcal{H}}\left[\int_0^t P[dX_u = 1|\mathcal{H}_u]\right]$$

Hence, based on the algebraic expression for $\widehat{E^X}(X_t)$ derived in equation (2.1.2),

$$\widehat{E^X(X_t^2)} \leq J_{t_{N_t}} \sum_{j=0}^{J_{t_{N_t}}} \{\sum_{t_i=t_1>0}^{t_i=t_{N_t}\leq t} \hat{P}(\Delta X_{t_i} = 1|H_{t_i}: N = j\ only)\hat{P}[H_{t_i}: N = j\ only]\}$$

And, $\widehat{\sigma_t^2} = \widehat{Var^X}(X_t) = \widehat{E^X(X_t^2)} - \{\widehat{E^X(X_t)}\}^2$

$$\leq J_{t_{N_t}} \sum_{j=0}^{J_{t_{N_t}}} \{\sum_{t_i=t_1>0}^{t_i=t_{N_t}\leq t} \hat{P}(\Delta X_{t_i} = 1|H_{t_i}: N = j\ only)\hat{P}[H_{t_i}: N = j\ only]\}$$

$$- \left[\sum_{j=0}^{J_{t_{N_t}}}\{\sum_{t_i=t_1>0}^{t_i=t_{N_t}\leq t} \hat{P}(\Delta X_{t_i} = 1|H_{t_i}: N = j\ only)\hat{P}[H_{t_i}: N = j\ only]\}\right]^2$$

Equivalently, $\widehat{Var^X}(X_t) \leq$

$$\left[\sum_{j=0}^{J_{t_{N_t}}}\{\sum_{t_i=t_1>0}^{t_i=t_{N_t}\leq t} \hat{P}(\Delta X_{t_i} = 1|H_{t_i}: N = j\ only)\hat{P}[H_{t_i}: N = j\ only]\}\right]\left[J_{t_{N_t}} - \right.$$

$$\left.\sum_{j=0}^{J_{t_{N_t}}}\{\sum_{t_i=t_1>0}^{t_i=t_{N_t}\leq t} \hat{P}(\Delta X_{t_i} = 1|H_{t_i}: N = j\ only)\hat{P}[H_{t_i}: N = j\ only]\}\right]$$

Hence, $\widehat{\sigma_t^2} \leq \hat{\mu}_t[J_{t_{N_t}} - \hat{\mu}_t] = \hat{\mu}_t[(x_t^{obs}) - \hat{\mu}_t]$

(2.2.1)

## 2.3         Asymptotic Theory

The Central Limit Theorem (CLT) for correlated variables (Chung [16], p 214) may be adapted to establish the asymptotic Normality of the sum of all events occurred to a subject in the following way.

By definition $X_t = \sum_{t_i=t_1>0}^{t_i=t_{N_t}\leq t} \Delta X_{t_i}$. The mean and variance of $X_t$ are defined as $\mu_t = E^X[X_t]$ and $\sigma_t^2 = Var^X[X_t]$.

Since $\{\Delta X_{t_i}, i \geq 1\}$ are dependent on $\{\mathcal{H}_{t_i}, i \geq 1\}$, if $\{\Delta X_{t_i}, i \geq 1\}$ are looked upon as a sequence of *uniformly bounded* random variables and let $\mathcal{F}_{t_i}$ be the Borel field generated by $\{\Delta X_{t_v}, 1 \leq v \leq i\}$, then the sequence can be called $r - dependent$ if and only if there exists an integer r such that for every i and h $\geq r + 1$, $\Delta X_{t_{i+h}}$ is independent of $\mathcal{F}_{t_i}$.

Given the above consideration, if $X_t = x_t^{obs}$ at T = t, (equivalent to saying $(X_t|\mathcal{H}_t) = x_t^{obs}$) and $\sigma_t$ is approximated by $\sqrt{\max(\sigma_t^2)}$, where the maximum value of $\sigma_t^2$ can be estimated as $\hat{\mu}_t\left[[(x_t^{obs}) - \hat{\mu}_t]\right]$, as shown in equation (2.2.1) (and $\mu_t$ can be estimated as presented in equation (2.1.2)), then $\frac{\sigma_t}{\sqrt[3]{x_t^{obs}}} \to \infty$ [which boils down to $\frac{\sqrt{x_t^{obs}}^2}{\sqrt[3]{x_t^{obs}}} \to \infty$ (assuming $\hat{\mu}_t$ being finite)] implies $\frac{X_t - \mu_t}{\sigma_t} \xrightarrow{a} N(0,1)$.

In a population consisting of n subjects, if $X_t^k$ denotes the number of events occurred to the $k^{th}$ individual by time t, then under the *I.I.D.* (independent and identically distributed) assumption for $X_t^1, X_t^2, \ldots, X_t^n$, $E^X\left(\frac{1}{n}\sum_{k=1}^{n} X_t^k\right) = \mu_t$, where $\mu_t$ is defined as $E^X(X_t^k) = \mu_t, \forall k = 1(1)n$.

Likewise, $Var^X\left(\frac{1}{n}\sum_{k=1}^{n} X_t^k\right) = \frac{\sigma_t^2}{n}$, where $\sigma_t^2$ is defined as $Var^X(X_t^k) = \sigma_t^2, \forall k = 1(1)n$; and based on equation (2.2.1), $\frac{\sigma_t}{\sqrt{n}} \leq \frac{\sqrt{\mu_t[\{Inf_{k=1(1)n}(x_t^{k,obs})\} - \mu_t]}}{\sqrt{n}}$.

Therefore, if the incidence rate of event occurrences in a population of n subjects by time t is denoted by $\frac{\sum_{k=1}^{n} X_t^k}{n}$, then under asymptotic normality of $X_t^k, \forall k = 1(1)n$ as shown above, the asymptotic sampling distribution of $\frac{1}{n}\sum_{k=1}^{n} X_t^k$ is $N\left(\mu_t, \frac{\sigma_t^2}{n}\right)$ when $X_t^k \to \infty, \forall k = 1(1)n$; or, to be precise, when $\frac{\sigma_t}{\sqrt[3]{x_t^{k,obs}}} \to \infty, \forall k = 1(1)n$. That is, $\frac{1}{n}\sum_{k=1}^{n} X_t^k \xrightarrow{a} N(\mu_t, \frac{\sigma_t^2}{n})$.

Alternatively, if $n \to \infty$, then also by Central Limit Theorem, $\frac{1}{n}\sum_{k=1}^{n} X_t^k \xrightarrow{a} N(\mu_t, \frac{\sigma_t^2}{n})$.

# 3 Data analysis

## 3.1 Homogeneous Poisson process (where event occurrences are independent of history of the process) without drop-out

100 Simulated datasets under different scenarios are created, each time with 100 simulated subjects having up to two events based on an exponential distribution with λ=0.003 and without any drop-out. If events did not occur in the first 370 days, or occurred after 370 days, then those respective data-points were truncated at day-370 to simulate a study-completion.

As presented in figure 3.1 below, in case of homogeneous Poisson process with no drop-outs, estimates based on the proposed estimator of the mean function are consistently close to the estimates based on the unbiased Nelson-Aalen estimate for the mean function.

**Figure 3.1** Scatter-plot of Nelson-Aalen estimate vs. proposed mean for homogeneous Poisson process without drop-out

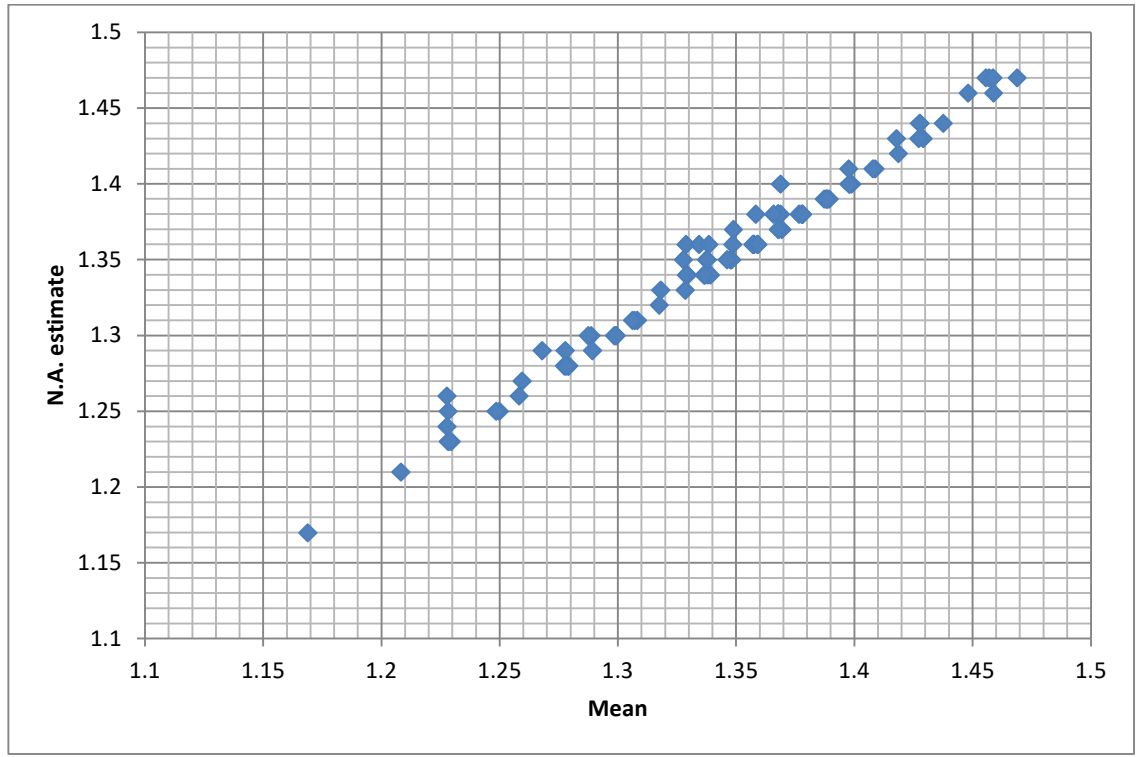

It is observed that here in this simulation study example, both the estimators of mean function (N.A. as well as the proposed estimator) over-estimate the theoretical mean function, the value of which was set at 1.11 (= 0.003*370 = $\rho t$) under assumption of a homogeneous Poisson process; which could be due to choosing a shorter time duration (370 days) for already converging to a Poisson distribution. In other words, had the time been truncated at a much larger or later time-point compared to day-370, the estimates could have converged to the theoretical mean.

### 3.2    Processes with event-dependednt intensity and drop-out

In this section, 100 simulations were run, where each simulation represented a population of 100 subject, each having up to two events. Here, along with the Exponential drop-out parameter ($\lambda_d = 0.001$), two different exponential parameters were considered for the event occurrence rates ($\lambda$=0.002 for the first event occurrence and $\lambda$=0.001 for the second event occurrence) to simulate dependence of the intensity of the occurrence of the second event on the first event.

In case of drop outs and intensities of occurrence of subsequent events being dependent on preceding events, estimates based on the proposed estimator of the mean function are most of the times consistently less than the corresponding estimates based on the Nelson-Aalen estimate for the mean function, as presented in figure 3.2. At this point, it should be noted that Nelson-Aalen estimate may not be an unbiased estimator of mean function when the intencity of event occurrence changes every time an event occurs, which is the scenario created here.

**Figure 3.2** **Scatter-plot of Nelson-Aalen estimate vs. proposed mean estimate for process with drop-outs and with intensity function of successive event occurrence changing with the occurrence of preceding event**

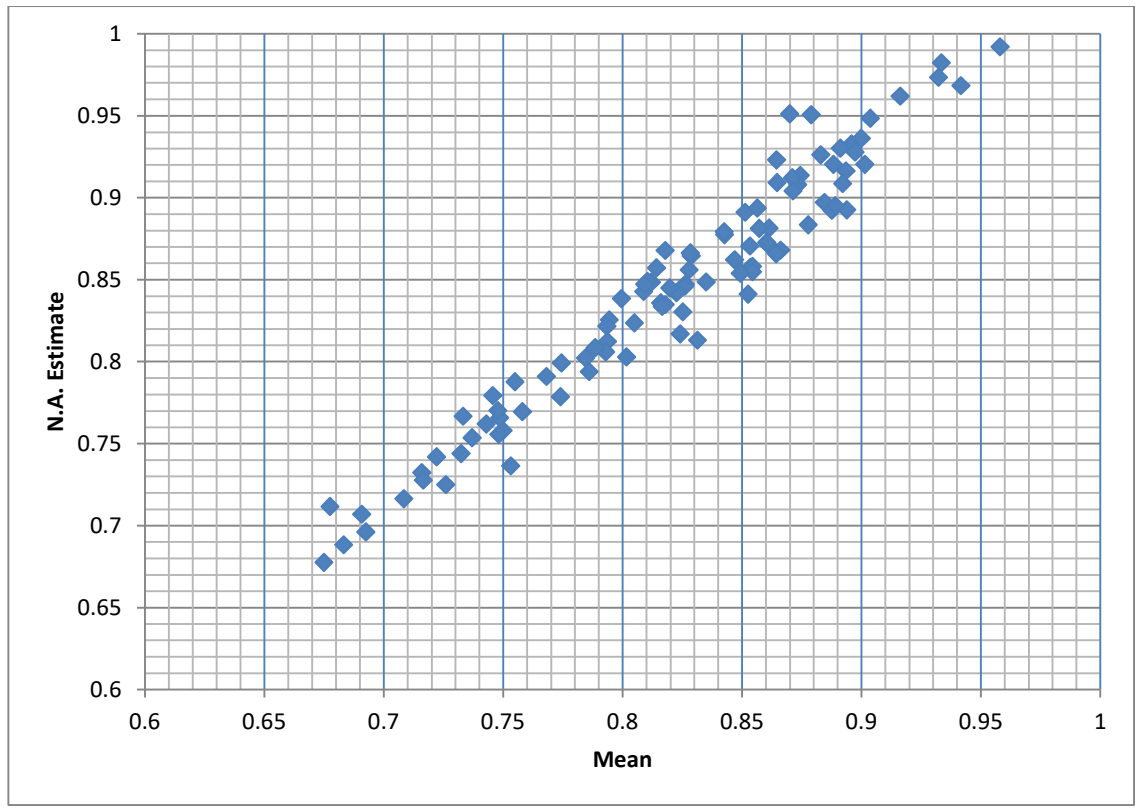

4   Discussion

The expected value of recurrent events specifies the central tendency and the first-order moment associated with the probability distribution of recurrent events at any given time-point. In Section 2 on mathematical development, the estimator for generalized mean function over time of event counts, the maximum value of variance over time for event counts, and also the asymptotic properties of event counts over time and mean event counts over time are derived for an event generating process, where occurrence of successive events depends on history of the process through preceding events. Therefore, the confidence interval of the estimate of mean counts can now be calculated for the above mentioned process, based on the asymptotic distribution at any given time-point of the estimator of mean function of event counts over time. In following sub-section, calculation of confidence interval and other application of the estimates are proposed through the use of "Incidence Rate", which is a widely used estimator of event occurrences in clinical and epidemiological studies.

## 4.1    Application of estimator of mean and maximum value of variance for a general counting process

In Section 2 on mathematical development, a non-parametric estimator for the mean function over time is derived for recurrent events following a counting process, where occurrence of successive events depends on history through preceding events. Although the variance of recurrent events could not be estimated, at least the maximum value of the variance over time could be estimated. In addition, when the total number of events occurred to each subject in a population or the number of subjects in that population is sufficiently large, then the incidence rate of event occurrences ($\frac{\sum_{k=1}^{n} X_t^k}{n}$) in that population by time $t$ is shown (in section 2.3) to follow asymptotic Normality. The mean of that asymptotic sampling distribution, $E(X_t)$ may be estimated by the algebraic expression presented in equation (2.1.2) (which might be a better estimate for mean counts than the crude estimate of the mean function by the incidence rate: $\frac{\sum_{k=1}^{n} X_t^k}{n}$) and the standard deviation can be approximated by $\frac{1}{\sqrt{n}}\sqrt{maximum\ value\ of\ the\ varience(X_t)}$, where the maximum value of the $Varience(X_t)$ is as presented by equation (2.2.1).

The property of asymptotic normality of incidence rate of recurrent events from a process where intensity of event occurrence changes with the occurrence of each higher order event may be used for calculating the asymptotic confidence interval of the estimated mean counts and may also be utilized especially in statistical Hypotheses testing, for testing the difference between treatment effects in clinical trials. Clearly, the length of confidence interval for the estimate of incidence rate of event occurrences in any population will be affected by the value of $\frac{\sqrt{smallest(x_t^{obs})}}{\sqrt{n}}$. In other words, if $\sqrt{Inf_{k=1(1)n}\{x_t^{k,obs}\}} \ll \sqrt{n}$, then the confidence interval will present an estimate of the mean of event occurrences with less margin of error.

The present work can be further extended to derive the estimator of mean count of events under informative censoring, or to derive the exact estimator of variance of event counts, for recurrent processes, where intensity of event occurrence changes with the occurrence of each higher order event.

**Acknowledgements:** The author thanks his former colleagues at Boehringer Ingelheim, especially Dr. Frank Fleischer and Dr. Erich Bluhmki for their motivating and valuable input over time in developing the concepts for dealing with the specific problem considered here.

sample study, Annals of Stat. 10 (1982) 1100-1120.

[11]  D.Y. Lin, L.J. Wei, I. Yang, Z. Ying, Semiparametric regression for the mean and rate function of recurrent events, J. of Royal Stat. Soc. B 62 (part 4) (2000) 711-730.

[12]  D.Y. Lin, L.J. Wei, Z. Ying, Semiparametric transformation models for point processes, J.of the Am. Stat. Assoc.96 (2001) 620-628.

[13]  M. Miloslavsky, S. Keles, M.J. van der Laan, S. Butler, Recurrent events analysis in the presence of time-dependent covariates and dependent censoring, J. of Royal Stat. Soc. B 66 (part 1) (2004) 239-257.

[14]  O.O Aalen, S. Johansen, An empirical transition matrix for non-homogeneous Markov chains based on censred observations, Scand. J.of Stat. 5 (1978) 141-150.

[15]  S.S. Menjoge, On Estimation of Frequency Data with Censored Observations. Pharma. Stat. 2 (2003) 191-197.

[16]  K.L Chung, A course in probability theory, 2nd ed, Academic Press, Inc., New York, 1974 p. 214.

21